\documentclass[aps,twocolumn,showpacs,superscriptaddress]{revtex4}

\usepackage{graphics}      
\usepackage{longtable}     
\usepackage{url}           
\usepackage{bm}            
\usepackage{graphicx}  
\usepackage{dcolumn}   
\usepackage{amssymb}   
\usepackage{amsmath}   
\usepackage{color}
\newcommand{\bea}{\begin{eqnarray}}
\newcommand{\eea}{\end{eqnarray}}
\newcommand{\be}{\begin{equation}}
\newcommand{\ee}{\end{equation}}
\newcommand{\ket}[1]{|#1\rangle}

\newcommand{\ave}[1]{\langle #1\rangle}
\newcommand{\lr}[1]{\left( #1\right)}

\newcommand{\nn}{\nonumber}

\begin{document}

\title{Optoelectromechanical transducer: reversible conversion between microwave and optical photons}
\author{Lin Tian\footnote{E-mail:~\textsf{ltian@ucmerced.edu}}}
\address{School of Natural Sciences, University of California, Merced, California 95343, USA}
\begin{abstract}
Quantum states encoded in microwave photons or qubits can be effectively manipulated, whereas optical photons can be coherently transferred via optical fibre and waveguide. The reversible conversion of quantum states between microwave and optical photons will hence enable the distribution of quantum information over long distance and significantly improve the scalability of hybrid quantum systems. Owning to technological advances, mechanical resonators couple to quantum devices in distinctly different spectral range with tunable coupling, and can serve as a powerful interface to connect those devices. In this review, we summarize recent theory and experimental progress in the coherent conversion between microwave and optical fields via optoelectromechanical transducers. The challenges and perspectives in achieving single-photon-level quantum state conversion will also be discussed.
\end{abstract}
\maketitle

\section{Overview}
Numerous quantum systems in the microwave domain, such as superconducting qubits and resonators, electron spins of nitrogen-vacancy center in diamond, and hyperfine states in ion trap qubits, have been studied for quantum information processing~\cite{SQ1, NVQbit1, IonQbit1}. The controllability of such systems enables the generation and manipulation of nonclassical states~\cite{SQbitR1, SQbitR2}. Meanwhile, optical photons can be easily distributed between distant nodes in a quantum network via optical fibre and waveguide~\cite{PhotonQbit1,Qnetwork1, Qnetwork2}. To facilitate the transfer of quantum information and to improve the scalability of the quantum processors, it would be desirable to build quantum interfaces that convert quantum states between microwave and optical photons. In such hybrid systems~\cite{HybridSystemReview}, the microwave photons and qubits form local information-processing units, while the optical photons play the role of a quantum data bus to transfer the information. In previous works~\cite{TransducerOptics1, TransducerOptics2, TransducerOptics3}, frequency conversion was demonstrated in the telecommunications wavelength, but its realization in the microwave-to-optical domain has only achieved very low efficiency~\cite{ElectroOptics1}. 

Mechanical resonators are macroscopic devices featuring collective lattice motion~\cite{OptomechanicsReview1, OptomechanicsReview2, OptomechanicsReview3, OptomechanicsReview4}. Due to the ubiquitous nature of the mechanical motion, such resonators couple with many kinds of quantum devices, from solid-state electronic circuits to atomic systems~\cite{MechanicsHybrid1, MechanicsHybrid2, MechanicsHybrid3, MechanicsHybrid4, MechanicsHybrid5, MechanicsHybrid6, MechanicsHybrid7, MechanicsHybrid8, MechanicsHybrid9, MechanicsHybrid10, MechanicsHybrid11}. In recent years, mechanical resonators ranging from acoustic frequency to radio frequency have been studied in their quantum limit. The mechanical modes can be prepared in their quantum ground state by various methods, such as thermalization in a cryogenic environment and cavity-assisted backaction cooling~\cite{Groundstate1, Groundstate2, Groundstate3, Groundstate4, Groundstate5, Groundstate6, Groundstate7}. In cavity optomechanical and electromechanical systems, the resolved-sideband regime can now be reached with the mechanical frequency far surpassing the cavity bandwidth~\cite{ResolvedSideband1, ResolvedSideband2, ResolvedSideband3, strongcouplingExp1, strongcouplingExp2}. Mechanically-induced transparency due to the destructive interference of the probe pulse with the anti-Stokes field generated by the pump pulse~\cite{EITExp1, EITExp2, EITExp3, EITExp4} and normal-mode splitting in the strong coupling limit~\cite{strongcouplingExp3, strongcouplingExp4, strongcouplingExp5, strongcouplingExp6, strongcouplingExp7} were observed. More recently, the coherent transfer of itinerant optical and microwave photon states to mechanical modes was demonstrated in state-of-art experiments~\cite{TranferMechanicalMode1, TranferMechanicalMode2}. This experimental progress indicates that the mechanical resonators can be utilized as an ideal interface to bridge quantum systems in distinctively different frequency ranges~\cite{MechanicalInterface1, MechanicalInterface2, MechanicalInterface3, MechanicalInterface4, MechanicalInterface5, MechanicalInterface6}.

Optoelectromechanical quantum transducers contain one mechanical resonator coupling in part to a microwave device and in part to an optical device. We can view such a transducer as an optomechanical system and an electromechanical system with their mechanical components merging into one. It was demonstrated that feedback cooling of the mechanical mode can be achieved in a cavity optoelectromechanical system by combining electrical actuation and optomechanical transduction~\cite{MicrowaveOpticalCooling}. In a commonly studied model, the mechanical resonator couples to a microwave cavity and an optical cavity via radiation-pressure interaction, with both cavities driven on their corresponding red sidebands. The effective coupling between each cavity and the mechanical mode is a beam-splitter-like interaction that can swap the states of these two modes. It was proposed that high-fidelity quantum state conversion between microwave and optical photons, either for intra-cavity states or for propagating photons, can be realized in an optoelectromechanical transducer via the excitation of the mechanical dark mode~\cite{MicrowaveOpticalTheo1, MicrowaveOpticalTheo2, MicrowaveOpticalTheo3, MicrowaveOpticalTheo4, MicrowaveOpticalTheo5, MicrowaveOpticalTheo6}. During the conversion, the effect of the mechanical noise could be significantly suppressed. In particular, the conversation efficiency for propagating photons in an ideal system reaches unity when the impedance matching condition is fulfilled. The first experimental observation of the mechanical dark mode was reported for two optical modes in a silica resonator~\cite{TransducerOptics3DarkMode}. An alternative microwave-to-optical transducer was proposed in an electro-optic system, where the interaction between the electrical and the optical modes resembles the optomechanical coupling~\cite{ElectroOptics2, ElectroOptics3}. At the moment, electro-optic modulators operate far from the regime for quantum state conversion~\cite{ElectroOptics1}, but improved performance on such devices could possibly be achieved to implement state conversion without mechanical resonators. 

Enormous progress has been achieved in converting microwave and optical photons via an optoelectromechanical transducer in several experimental setups. In one experiment, the transducer is constructed with a piezoelectric optomechanical crystal~\cite{MicrowaveOpticalExp1}.  A microwave signal in resonance with the mechanical mode is coherently converted to propagating photons in optical frequency. In another experiment, a radio-frequency field in resonance with an LC oscillator was converted to an optical signal reflected from the surface of a nanomembrane~\cite{MicrowaveOpticalExp2}. Both experiments were conducted at room temperature. Recently, bi-directional state conversion between microwave and optical photons was demonstrated with a microwave resonator and a Fabry P\'{e}rot cavity coupling to a membrane~\cite{MicrowaveOpticalExp3}. In this experiment, the amplitude information and the phase information of the photon states are preserved in the cavity output. 

The study of quantum state conversion via optoelectromechanical transducers is in its infancy. The demonstration of the coherent conversion between microwave and optical fields indicates that it is promising to convert photons in the quantum limit in the near future. To reach high conversion efficiency at the single-photon level, however, various technical challenges have to be overcome. One dominant factor that restricts the conversion efficiency is the thermal noise of the mechanical mode. Even when the impedance matching condition is satisfied~\cite{MechanicalInterface2, MechanicalInterface3, MechanicalInterface4, TransducerOptics3DarkMode}, the mechanical noise still exerts stringent requirement on the cooperativity and the operating temperature of the transducer. Another experimental hurdle is the intrinsic dissipation of a non-ideal cavity, which could seriously reduce the signal-to-noise ratio of the output states. It will require carefully designed experimental systems, accurate control over the external drivings, and quantum-limited measurement apparatus to reach high conversion efficiency~\cite{Detection1}. 

The optoelectromechanical interface can also be utilized for other quantum applications. By driving one cavity on the first blue sideband, effective coupling in the form of a parametric down-conversion operation is obtained, which is a key element in producing continuous-variable entanglement, mode amplification, and squeezing~\cite{Entanglement2Mode1, Entanglement2Mode2, Entanglement2Mode3, Entanglement3Mode1, Entanglement3Mode2, Amplification1, Amplification2}. When combined with the beam-splitter operation, this coupling induces strong entanglement between microwave and optical photons or photon-phonon pairs ~\cite{EntanglementInterface1, EntanglementInterface2, EntanglementInterface3, EntanglementInterface4, EntanglementInterface5}, which can be used to implement quantum teleportation protocols~\cite{TeleportationInterface1, TeleportationInterface2}. These applications make this quantum interface a preferable component in hybrid quantum networks. 

In this review, we will summarize the theory and experimental achievements in the study of converting microwave and optical photons via optoelectromechanical transducers. In Sec.~\ref{sec:interface}, we describe the basic quantum operations and the general framework for studying electromechanical and optomechanical interfaces. The quantum state conversion approaches for intra-cavity photons and itinerant photons are presented in Sec.~\ref{sec:conversion}. The conditions for achieving high conversion efficiency are also discussed in this section. Other quantum applications of optoelectromechanical interfaces, such as entanglement generation, are presented in Sec.~\ref{sec:otherapp}. We summarize recent experimental progress in Sec.~\ref{sec:experiments}, focusing on the state conversion between microwave and optical fields. Finally, the challenges and perspectives in achieving single-photon level quantum state conversion are discussed in Sec.~\ref{sec:discussions}. 

\section{Light-matter interface\label{sec:interface}}
\subsection{System\label{subset:coupling}}
We start with a typical electromechanical or optomechanical system containing a cavity mode coupling to a mechanical mode via radiation-pressure interaction. The coupling has the form $H_{int}=\hbar G \hat{a}^{\dag}\hat{a}(\hat{b}_{m}+\hat{b}_{m}^{\dag})$~\cite{OptomechanicsInteraction1}, where $\hat{a}$ ($\hat{b}_{m}$) is the annihilation operator of the cavity (mechanical) mode, and $G$ is the single-photon coupling strength. This interaction can be viewed as a shift of the cavity frequency $\omega_{c}$ by the mechanical vibration. In the resolved-sideband regime where the frequency of the mechanical mode $\omega_{m}$ exceeds the cavity bandwidth $\kappa$, the cavity spectrum is affected by this interaction to include sidebands at frequencies $\omega_{c}\pm n\omega_{m}$ with $n$ being an integer number. We can also regard this interaction as a force on the mechanical mode that depends on the cavity photon number. By applying a strong driving on the cavity mode, this interaction is linearized into tunable couplings that can be used to manipulate or probe the mechanical motion.  

\subsection{Linearized quantum operations\label{subsec:linearizing}}
The strength of the single-photon coupling $G$ is usually very weak~\cite{OptomechanicsReview4}. For example, the microwave device in Ref.~\cite{Groundstate3} has a mechanical frequency of $\omega_{m}/2\pi=10\,\textrm{MHz}$ and a cavity bandwidth of $\kappa/2\pi=170\,\textrm{kHz}$, whereas $G/2\pi=400\,\textrm{Hz}$. The light-matter interaction can be enhanced by driving the cavity mode strongly. The driving excites large number of photons in the cavity with photon number $n_{ph}\sim10^{6-8}$. The radiation-pressure interaction is then transformed into effective linear or bilinear couplings with coupling strength $g=G\sqrt{n_{ph}}$, which is increased by several orders of magnitude. In recent experiments, the strong coupling limit with $g\gg\kappa$ was realized in both microwave and optical cavities. Below we assume $g$ to be a real number for simplicity of discussion.  

\begin{figure}
\includegraphics[width=\columnwidth,clip]{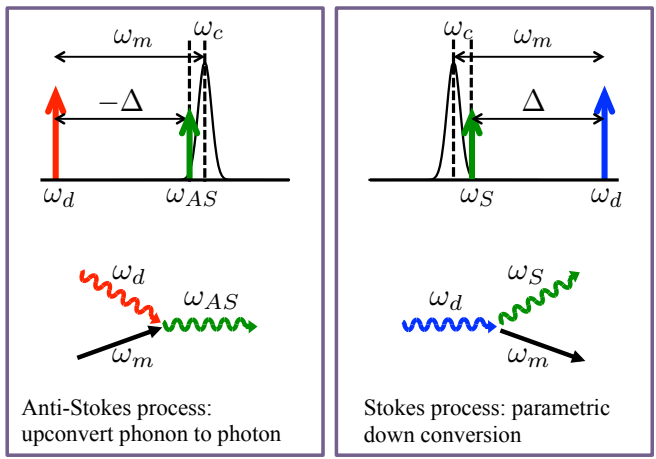}
\caption{Linearized operations. Left panel: anti-Stokes process described by the linear coupling in Eq.~(\ref{eq:Hl}); right panel: Stokes process described by the bilinear coupling in Eq.~(\ref{eq:Hbl}). The scattered photons are labelled by $\omega_{AS}$ and $\omega_{S}$.}
\label{fig1}
\end{figure}
The effective couplings strongly depend on the detuning of the driving field $\Delta=\omega_{d}-\omega_{c}$ with $\omega_{d}$ being the driving frequency. For $\Delta=-\omega_{m}$ with the cavity driven on the first red sideband of the mechanical mode, the effective coupling is dominated by a beam-splitter-like interaction. In the interaction picture, the Hamiltonian is~\cite{Linearization1}
\be
H_{I}=\hbar g(\hat{a}^{\dagger}\hat{b}_{m}+\hat{b}_{m}^{\dagger}\hat{a}),\label{eq:Hl}
\ee
which describes an anti-Stokes process. An incoming photon with frequency $\omega_{d}$ is scattered into the cavity resonance by absorbing a phonon of the mechanical mode, as illustrated in the left panel of Fig.~\ref{fig1}. The driving field serves as a parametric source that up-converts a phonon to a cavity photon. In the interaction picture,  the system operators at time $t$ evolve as 
\begin{subequations}
\begin{align}
&\hat{a}(t)=\cos(gt)\hat{a}(0)-i\sin(gt)\hat{b}_{m}(0),\label{eq:evolHl1}\\
&\hat{b}_{m}(t)=\cos(gt)\hat{b}_{m}(0)-i\sin(gt)\hat{a}(0),\label{eq:evolHl2}
\end{align}
\end{subequations}
written in terms of the operators at $t=0$. This is exactly the beam-splitter operation in quantum optics~\cite{InputOutput}. At time $t=\pi/2g$, $\hat{a}(t)=-i\hat{b}(0)$ and $\hat{b}(t)=-i\hat{a}(0)$. The operators are exchanged with each other up to the phase factor $-i$. In the Schr\"{o}dinger picture, this is equivalent to swapping the states of the cavity and the mechanical modes. This operation is at the heart of quantum state conversion as well as cavity cooling of the mechanical mode. It can also be used to detect the mechanical state by converting the state to an output photon. 

The beam-splitter-like interaction generates discrete-state entanglement. Applying this operation for a duration of $t=\pi/4g$ on the initial state $\ket{1_{c}0_{m}}$, the final state becomes $\lr{\ket{1_{c}0_{m}}-i\ket{0_{c}1_{m}}}/\sqrt{2}$, where $\ket{n_{c}n_{m}}$ denotes a Fock state with $n_{c}$ photons and $n_{m}$ phonons. The final state is a maximally entangled state between the cavity and the mechanical modes in the Fock-state basis. 

When the cavity is driven on the first blue sideband with $\Delta=\omega_{m}$, the dominant interaction is a parameter-down-conversion operation with
\be
H_{I}=i \hbar g(\hat{a}^{\dagger}\hat{b}_{m}^{\dag}-\hat{b}_{m}\hat{a}).\label{eq:Hbl}
\ee
This interaction describes a Stokes process with an incoming photon scattered into the cavity resonance and simultaneously emitting a phonon, as shown in the right panel of Fig.~\ref{fig1}. The operators evolve as
\begin{subequations}
\begin{align}
&\hat{a}(t)=\cosh(gt)\hat{a}(0)+\sinh(gt)\hat{b}_{m}^{\dag}(0),\label{eq:evolHbl1}\\
&\hat{b}_{m}^{\dag}(t)=\cosh(gt)\hat{b}_{m}^{\dag}(0)+\sinh(gt)\hat{a}(0),\label{eq:evolHbl2}
\end{align}
\end{subequations}
which corresponds to a Bogoliubov transformation between these two modes. This operation generates two-mode squeezing and continuous-variable entanglement ~\cite{ContinuousVariableRMP}. Various quantum protocols such as quantum teleportation of an unknown state can be implemented with this entanglement.  

The linear and bilinear couplings discussed above have the advantage of being easily tunable. By varying the amplitude of the driving field, the average photon number $n_{ph}$, and hence the coupling strength $g$, is adjustable in a broad range.

The radiation-pressure interaction in Sec.~\ref{subset:coupling} contains only the adiabatic term in the light-matter coupling~\cite{OptomechanicsInteraction1}. In a multimode system, the non-adiabatic terms also play an important role in controlling the system. One such term is $H_{int} = \hbar G \hat{a}_{1}^{\dag}\hat{a}_{2} (\hat{b}_{m}+\hat{b}_{m}^{\dag})+h.c.$~\cite{OptomechanicsInteraction2}, where the interaction is between two cavity modes $\hat{a}_{1},\,\hat{a}_{2}$, and a mechanical mode $\hat{b}_{m}$. As shown in Ref.~\cite{MechanicalInterface5}, by driving one of the cavity modes into classical motion, a parametric coupling between the other cavity and the mechanical modes is created. This parametric coupling has the form of Eq.~(\ref{eq:Hl}) or Eq.~(\ref{eq:Hbl}), depending on the driving.

Other forms of light-matter interaction can be obtained by varying the circuit geometry of the system. In a membrane-in-the-middle setup~\cite{strongcouplingExp1, EITExp4}, the coupling is varied from a linear to a quadratic dependence on the mechanical motion by adjusting the position of the membrane inside the optical cavity with $H_{int}=\hbar G \hat{a}^{\dag} \hat{a} \hat{b}_{m}^{\dag} \hat{b}_{m}$. Using such a coupling, a quantum nondemolition measurement of the mechanical mode can be conducted by detecting the optical field. 

Recently, the single-photon strong (ultra-strong) coupling limit with $G\sim \omega_{m}$ was studied in a number of theoretical works~\cite{Ultrastrong1, Ultrastrong2, Ultrastrong3, Ultrastrong4}. In this limit, the intrinsic nonlinearity of the coupling induces novel quantum effects such as photon blockade and non-Gaussian nature of the quantum states. 

\subsection{Heisenberg-Langevin equation\label{subsec:langevin}}
The electromechanical and optomechanical systems are subject to environmental fluctuations that induce damping and noise in the evolution of the system operators. The dissipation of the cavity mode includes intrinsic damping and leakage to the external channels. The latter generates output field at the end mirror of an optical cavity or transmission from a microwave resonator to an open transmission line. For the mechanical mode, the dissipation is associated with its finite quality factor when it is not directly driven by an external field. Below we incorporate the contribution of the fluctuations into the equations of motion for the system operators.

We assume that the total damping rate of the cavity is $\kappa=\kappa_{ext}+\kappa_{in}$, including the transmission rate to the external circuit $\kappa_{ext}$ and the intrinsic damping rate $\kappa_{in}$. The total input operator is $\hat{a}_{in}(t)=\sqrt{\kappa_{ext}/\kappa}\hat{a}_{p,in}(t)+\sqrt{\kappa_{in}/\kappa}\hat{a}_{s,in}(t)$ with an incoming photon operator $\hat{a}_{p,in}$ (the probe field) and an intrinsic noise operator $\hat{a}_{s,in}(t)$. The damping rate of the mechanical mode is $\gamma_{m}$ with a noise operator $\hat{b}_{in}(t)$. We let the noise operators obey the Markovian correlation functions: $\ave{\hat{a}_{s,in}(t)\hat{a}_{s,in}^{\dag}(t^{\prime})} = (n_{th}^{(c)} + 1)\delta(t-t^{\prime})$ and $\ave{\hat{b}_{in}(t)\hat{b}_{in}^{\dag}(t^{\prime})}=(n_{th}^{(m)}+1)\delta(t-t^{\prime})$, with $n_{th}^{(c)}$ ($n_{th}^{(m)}$) being the thermal occupation number of the cavity (mechanical) mode. For an optical cavity, $n_{th}^{(c)}\approx 0$. The Heisenberg-Langevin equations for the system operators are~\cite{Linearization1}
\begin{subequations}
\begin{align}
&\dot{\hat{a}}=i\Delta\hat{a}-ig\lr{\hat{b}_{m}+\hat{b}_{m}^{\dag}}-\frac{\kappa}{2}\hat{a}+\sqrt{\kappa}\hat{a}_{in}(t);\label{eq:langevina}\\
&\dot{\hat{b}}_{m}=-i\omega_{m}\hat{b}_{m}-ig\lr{\hat{a}+\hat{a}^{\dag}}-\frac{\gamma_{m}}{2}\hat{b}_{m}+\sqrt{\gamma_{m}}\hat{b}_{in}(t).\label{eq:langevinb}
\end{align}
\end{subequations}
Following these equations, the system operators can be written as integral functions of the input operators. The cavity output operator in the external circuit $\hat{a}_{p,out}(t)$ is derived using the boundary condition of the cavity~\cite{InputOutput}. Based on the convention of the operator phases adopted in Ref.~\cite{MicrowaveOpticalTheo4}, we derive the input-output relation: $\hat{a}_{p,out}(t)=\hat{a}_{p,in}(t)-\sqrt{\kappa_{ext}}\hat{a}(t)$. 

We consider the system to be in the resolved-sideband regime. In the strong coupling limit with $g\gg\kappa$, the dynamics of the system operators is dominated by the effective coupling in Eqns.~(\ref{eq:Hl}, \ref{eq:Hbl}), while the damping and noise terms play a secondary role. For $\Delta=-\omega_{m}$, the conversion between the intra-cavity state of $\hat{a}$ and the mechanical state can be realized. In contrast, in the weak coupling limit with $g\ll\kappa$, when $\kappa_{ext}\gg\kappa_{in}$, an itinerant photon state injected into the cavity can be reversibly converted to a mechanical state. To illustrate this process, we adiabatically eliminate the cavity mode from Eqns.~(\ref{eq:langevina}, \ref{eq:langevinb}) and apply the rotating-wave approximation (RWA). The dynamics of the mechanical mode is then governed by 
\be
\dot{\hat{b}}_{m}=i\sqrt{\Gamma}\hat{a}_{in}(t)-\frac{\Gamma+\gamma_{m}}{2}\hat{b}_{m}+\sqrt{\gamma_{m}}\hat{b}_{in}(t)\label{eq:mconvert}
\ee
with $\Gamma=4g^{2}/\kappa$. This equation describes the conversion between the cavity input $\hat{a}_{in}(t)$ and the mechanical mode. Let the main component of the input operator $\hat{a}_{in}(t)$ be the incoming field $\hat{a}_{s,in}(t)$. Itinerant photon states can hence be transferred to (and from) the mechanical mode via this interface. Without an incoming field, the vacuum state is continuously converted to the mechanical mode, which exactly corresponds to the cavity cooling process with the cooling rate $\Gamma$~\cite{OptomechanicsReview1}. Note that the heating (counter-rotating) terms are omitted from Eq.~(\ref{eq:mconvert}) under the RWA. 

The Heisenberg-Langevin equations can be written in the frequency domain when the effective coupling is time-independent. The Fourier transformation of an arbitrary operator $\hat{o}(t)$ is $\hat{o}(t)=\int {d\omega/\sqrt{2\pi}} e^{-i\omega t}\hat{o}(\omega)$ with the frequency component $\hat{o}(\omega)$. In the frequency domain, Eqns.~(\ref{eq:langevina},\ref{eq:langevinb}) become
\begin{subequations}
\begin{align}
&-i\varepsilon_{a}\hat{a}(\omega)+ig \lr{\hat{b}_{m}(\omega)+\hat{b}_{m}^{\dag}(-\omega)}=\sqrt{\kappa}\hat{a}_{in}(\omega),\label{eq:langevinfreqa}\\
&-i\varepsilon_{b}\hat{b}(\omega)+ig \lr{\hat{a}(\omega)+\hat{a}^{\dag}(-\omega)}=\sqrt{\gamma_{m}}\hat{b}_{in}(\omega),\label{eq:langevinfreqb}
\end{align}
\end{subequations}
with $\varepsilon_{a}=\omega+\Delta+i(\kappa/2)$ and $\varepsilon_{b}=\omega-\omega_{m}+i(\gamma_{m}/2)$. Together with their conjugate equations (not shown), we obtain a set of four equations for the frequency components $\{\hat{a}(\omega),\,\hat{b}(\omega),\,\hat{a}^{\dag}(-\omega),\,\hat{b}^{\dag}(-\omega)\}$. The system operators and the output operators can be derived by solving these linear equations in terms of the input operators. 

\subsection{Coherent effects\label{subsec:coherenceeffect}}
The linear coupling in Eq.~(\ref{eq:evolHl1}) is at the root of several coherent effects observed in mechanical systems. The mechanically-induced transparency was demonstrated in microwave and optical cavities when the pump field is red-detuned with $\Delta=-\omega_{m}$~\cite{EITTheo1, EITTheo2}. In the weak coupling limit with $g\ll\kappa$, the system is described as coupled harmonic oscillators subject to heavy damping. In the experiments, a probe pulse near the cavity resonance is injected to the cavity. The transmission spectrum of the probe field contains a sharp peak at the cavity resonance $\omega=\omega_{c}$ due to the destructive interference between the probe field and the anti-Stokes field generated by the pump pulse. The peak corresponds to a narrow window of high transmission rate. This effect is similar to the electromagnetically-induced transparency that is well studied in atomic systems~\cite{EITAtom}. Increasing the pump power to reach the strong coupling limit with $g\gg\kappa$, the transmission spectrum of the probe field splits into two dips at the frequencies $\sim(\omega_{c}\pm g)$, respectively, due to the normal-mode splitting~\cite{strongcouplingExp3, strongcouplingExp4, strongcouplingExp5}. These dips correspond to the excitation of the normal modes in the linearized system $(\hat{a}\pm\hat{b})/\sqrt{2}$, which are polariton-like modes composed of both the cavity and the mechanical modes. More complicated setups were studied for related effects such as the mechanically-induced amplification~\cite{EITTheo3}. These effects indicate the coherent nature of the coupling between the cavity and the mechanical modes. 

\section{Quantum state conversion\label{sec:conversion}}
\begin{figure}
\includegraphics[width=\columnwidth,clip]{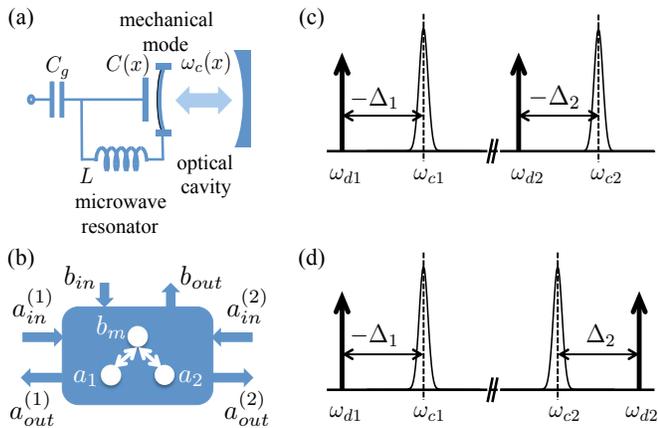}%
\caption{Microwave-to-optical interface. (a) A schematic circuit of a transducer made of a microwave resonator and an optical cavity connected to a mechanical resonator. The electromechanical (optomechanical) coupling is represented by the dependence of the capacitance of the microwave cavity $C(x)$ (the optical frequency $\omega_{c}(x)$) on the mechanical displacement $x$, with $L$ the inductance and $C_{g}$ a gate capacitance of the microwave cavity. (b) The input and and output ports of the cavity and the mechanical modes. (c) and (d) The spectrum of the cavity modes and the driving fields for the state conversion with $\Delta_{1} = \Delta_{2} = -\omega_{m}$ and the entanglement generation with $\Delta_{1} = -\Delta_{2} =-\omega_{m}$, respectively. The cavity frequencies are $\omega_{ci}$ and the driving frequencies are $\omega_{di}$ with $i=1,\,2$.}
\label{fig2}
\end{figure}

\subsection{The transducer\label{subsec:transducer}}
An optoelectromechanical transducer can be formed by merging an electromechanical system with an optomechanical system~\cite{OptomechanicsReview1, MicrowaveOpticalCooling}. The mechanical resonator in the transducer acts as a bridge that reversibly converts information between the microwave and the optical devices. In this review, we focus on a widely studied setup for the transducer: a microwave cavity and an optical cavity both coupling to a mechanical resonator, as illustrated in Fig.~\ref{fig2} (a, b)~\cite{MicrowaveOpticalTheo1, MicrowaveOpticalTheo2, MicrowaveOpticalTheo3, MicrowaveOpticalTheo4, MicrowaveOpticalTheo5}. A number of variations of this setup were also studied for the state conversion. For example, the mechanical resonator can couple directly to an electrical driving field via piezoelectric actuation or to a propagating optical field by surface reflection~\cite{MicrowaveOpticalExp1, MicrowaveOpticalExp2}. 

Denote the cavity modes as $\hat{a}_{1}$ and $\hat{a}_{2}$. Each cavity is driven by a red-detuned field with the detuning $\Delta_{i}$ ($i=1,\,2$). The spectrum of the cavity modes together with the driving fields is shown in Fig.~\ref{fig2} (c) for the state conversion protocols. The electromechanical and optomechanical couplings are linearized into beam-splitter-like interaction with $H_{I}=\sum_{i=1,2}\hbar g_{i}(t)(\hat{a}_{i}^{\dagger}\hat{b}_{m}+\hat{b}_{m}^{\dagger}\hat{a}_{i})$. The coupling strengths $g_{i}(t)$ are tunable by varying the driving fields. Taking damping and noise into account, the Heisenberg-Langevin equations for the modes in the transducer can be written as a vector equation~\cite{MicrowaveOpticalTheo2, MicrowaveOpticalTheo3}
\be
d\vec{v}(t)/dt=-i M(t)\vec{v}(t)+\sqrt{K}\vec{v}_{in}(t)\label{eq:transducerlangevin}
\ee
with $\vec{v}(t)=[\hat{a}_{1}(t),\hat{b}_{m}(t),\hat{a}_{2}(t)]^{\textrm{T}}$ for the system operators, $\vec{v}_{in}(t)=[\hat{a}_{in}^{(1)}(t),\hat{b}_{in}(t),\hat{a}_{in}^{(2)}(t)]^{\textrm{T}}$ for the input operators, the matrix  
\be
M(t)=\left(\begin{array}{ccc}
-\delta_{1}-i\frac{\kappa_{1}}{2} & g_{1}(t) & 0\\
g_{1}(t) & -i\frac{\gamma_{m}}{2} & g_{2}(t)\\
0 & g_{2}(t) & -\delta_{2}-i\frac{\kappa_{2}}{2}
\end{array}\right), \label{eq:Mt}
\ee
and the diagonal matrix $K=\textrm{diag}(\kappa_{1},\gamma_{m},\kappa_{2})$. The detuning $\delta_{i}$ is defined in the interaction picture with $\delta_{i}=\Delta_{i}+\omega_{m}$. The cavity damping rate includes both the intrinsic damping rate $\kappa_{i,in}$ and the leakage to the external circuit $\kappa_{i,ext}$ with $\kappa_{i}=\kappa_{i,in}+\kappa_{i,ext}$, and the cavity input operators are defined similarly.

\subsection{Conversion of intra-cavity states\label{subsec:intracavity}}
Within a single node of a quantum network, a qubit operated at the microwave frequency can directly interact with a microwave cavity and swap its state with the cavity mode. The conversion of the intra-cavity states between the microwave and the optical modes will hence facilitate the distribution of qubit states to a distant node. 

\begin{figure}
\includegraphics[width=\columnwidth,clip]{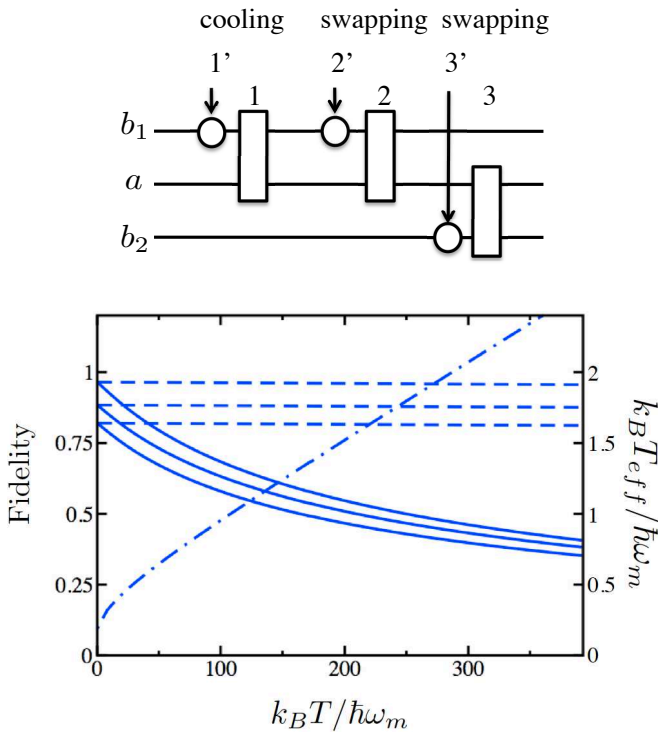}
\caption{Pulse sequence and fidelity of intra-cavity state conversion versus bath temperature~\cite{MechanicalInterface4}. Top: pulse sequence of the double-swap scheme plus the pre-cooling pulse. Bottom: conversion fidelity. The solid (dashed) lines: without (with) the pre-cooling pulse. The lines from top to bottom are for the Gaussian states $|1,\,0\rangle$, $|2,\,0\rangle$, and $|2,\,0.4\rangle$, respectively, with the states defined as $|\alpha, r_0\rangle= e^{\alpha a^\dag - \alpha^{*} a} e^{(r_0^{*}a^2-r_0 a^{\dag 2})/2}|0\rangle$. The dot-dashed line is the effective temperature $T_{eff}$ defined by the residue mechanical fluctuations after the cooling pulse. The parameters are $\omega_m=1$, $g_{1}=0.1$, $g_{2}=0.07$, $\kappa/\omega_m = 0.01$, and $\gamma_m/\omega_m=10^{-5}$, in arbitrary units. }
\label{fig3}
\end{figure}

In a simple approach, an initial state prepared in cavity $\hat{a}_{1}$ can be transferred to cavity $\hat{a}_{2}$ in two steps with the transducer operated in the strong coupling limit~\cite{MechanicalInterface2, MechanicalInterface4}. First, turn on the effective coupling $g_{1}$ (with $g_{1}\gg\kappa_{1}$) for a duration of $\pi/2g_{1}$. This operation swaps the initial cavity state to the mechanical mode. Then, turn on the effective coupling $g_{2}$ (with $g_{2}\gg\kappa_{2}$) for a duration of $\pi/2g_{2}$ to swap the state in the mechanical mode to the second cavity. Define the conversion fidelity as $F(\rho_{i},\rho_{f})=[\textrm{Tr}(\sqrt{\rho_{a1}}\rho_{a2}\sqrt{\rho_{a1}})^{1/2}]^{2}$ between the density matrices $\rho_{a1}$ of the initial state in $\hat{a}_{1}$ and $\rho_{a2}$ of the final state in $\hat{a}_{2}$. The conversion fidelity of this simple approach could be seriously degraded by the thermal noise in the mechanical mode. In particular, when $\gamma_{m}n_{th}^{(m)}>\kappa_{i}$, the mechanical noise becomes the dominant noise source. To improve the fidelity, a pre-cooling pulse, which also swaps the states of $\hat{a}_{1}$ and $\hat{b}_{m}$ but with the cavity prepared in the vacuum state, could be added before the two-step state conversion scheme~\cite{MechanicalInterface4}. This pulse swaps the vacuum state to the mechanical mode and can be viewed as a transient cooling process. The pre-cooling pulse hence removes the thermal noise and increases the conversion fidelity, as shown in Fig.~\ref{fig3}. This approach, however, is limited by the re-heating of the mechanical mode which occurs within a time interval of $1/\gamma_{m}n_{th}^{(m)}$.

To suppress the effect of the mechanical noise, an adiabatic approach based on the concept of the mechanical dark mode was proposed~\cite{MicrowaveOpticalTheo2, MicrowaveOpticalTheo3, MicrowaveOpticalTheo4}. This approach is similar to the adiabatic state transfer process in a $\Lambda$-system which converts atoms from one ground state to another. Under two-photon resonance with $\Delta_{1}=\Delta_{2}=-\omega_{m}$, the effective Hamiltonian $H_{I}$ has an eigenmode: $\psi_{d}=[-g_{2},0,g_{1}]^{\text{T}}/g_{0}$ at the given couplings, with $g_{0}=\sqrt{g_{1}^{2}+g_{2}^{2}}$. This mode is a mechanical ``dark'' mode, not containing any mechanical component. The other eigenmodes of $H_{I}$ are ``bright'' modes which are superpositions of the cavity and the mechanical modes. The eigenvalue of the dark mode is zero, well separated from the eigenvalues of the bright modes $\pm g_{0}$. In the strong coupling limit, as the coupling strengths vary adiabatically (following the Landau-Zener condition~\cite{LandauZener}), a quantum state initially stored in the dark mode will be preserved in the dark mode, and the perturbation of the mechanical noise will be greatly reduced. At the initial time of the state conversion, $g_{1}=0$ and $g_{2}$ is a finite negative value with the dark mode $\psi_{d}=\hat{a}_{1}$. Then $g_{1}$ is adiabatically increased to reach a finite value and $g_{2}$ is adiabatically switched to zero at the final time $T_{f}$ when the dark mode becomes $\psi_{d}=\hat{a}_{2}$. Hence an initial state encoded in $\hat{a}_{1}$ is transferred to $\hat{a}_{2}$ at time $T_{f}$. The mechanical mode is only virtually excited due to the destructive interference between the couplings. It was shown that the contribution of the mechanical noise in the conversion fidelity is reduced by a factor of $(\kappa_{1}-\kappa_{2})/4g_{0}]^{2}\ll 1$, making the state conversion robust against mechanical noise. 

During the adiabatic approach, the coupling strengths are varied slowly to meet the adiabatic condition. The state conversion is accomplished in a time scale much longer than $1/g_{0}$. This makes the approach more vulnerable to cavity dissipation. In the double-swap approach, the state conversion is realized in a shorter time scale of $1/g_{i}$, but is subject to severe perturbation of the mechanical noise. In Ref.~\cite{MicrowaveOpticalTheo4}, a hybrid conversion scheme that combines the ideas of these two approaches was studied. It was shown that the method that achieves optimal conversion fidelity depends on the ratio $\kappa_{i}/g_{i}$ and the temperature $T$. For example, the adiabatic approach yields optimal fidelity in the high temperature limit when the mechanical noise is strong. 

A Raman-like process with $\vert\Delta_{i}+\omega_{m}\vert\gg g_{i}$ and $\Delta_{1}=\Delta_{2}$ is also studied for high-fidelity intra-cavity state conversion~\cite{MicrowaveOpticalTheo3, MicrowaveOpticalTheo6}. Here the couplings $g_{i}$ generate a Rabi flip between the cavity modes with the Rabi frequency $g_{1}g_{2}/|\Delta_{i}+\omega_{m}|$. The large frequency offset $\left|\Delta_{i}+\omega_{m}\right|$ prevents the mechanical mode from being excited, which suppresses the effect of the mechanical noise. 

\subsection{Conversion of itinerant states\label{subsec:itinerant}}
\begin{figure}
\includegraphics[width=\columnwidth,clip]{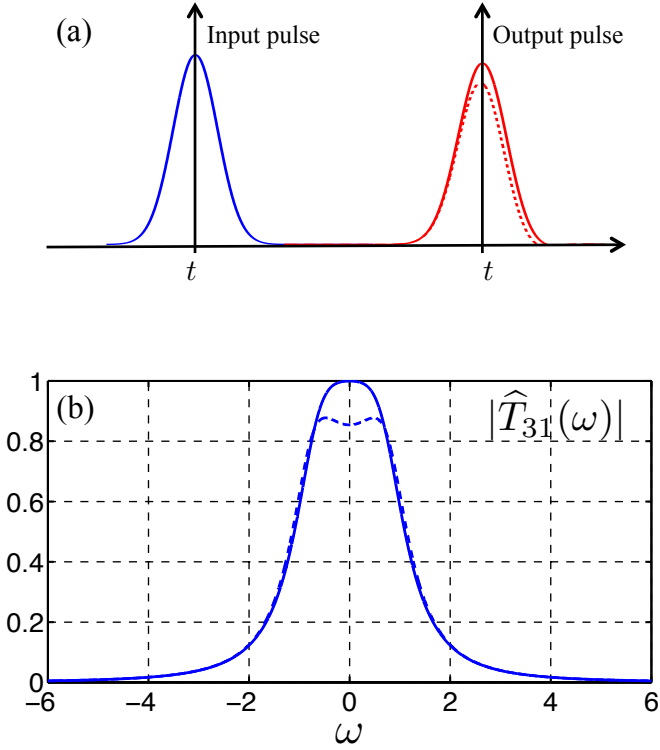}
\caption{Itinerant state conversion. (a) The operator average of the output field $\ave{\hat{a}_{p,out}^{(2)}(t)}/A$. The input field has Gaussian time dependence with $\ave{\hat{a}_{p,in}^{(1)}(t)}=Ae^{-\sigma_{p}^{2}t^{2}/2}$ and $\sigma_{p}=0.4$. (b) The conversion matrix element $\vert\widehat{T}_{31}(\omega)\vert$. In both (a) and (b), the solid lines are for $\kappa_{1}=3.2$ and $\kappa_{2}=1.8$ which satisfy the impedance matching condition; and the dashed lines are for $\kappa_{1}=1.8$ and $\kappa_{2}=3.2$. Other parameters are $g_{1}=0.8$, $g_{2}=0.6$, and $\gamma_{m}=0.001$. The conversion halfwidth is $\Delta\omega\sim\Gamma_{i}\sim1$. We assume no intrinsic dissipation and the parameters are in arbitrary units.}
\label{fig4}
\end{figure}

A propagating state can be converted via an optoelectromechanical interface. Consider an incoming state represented by the input operator $\hat{a}_{p,in}^{(1)}$. The total input operator of cavity $\hat{a}_{1}$ is $\hat{a}_{in}^{(1)}=\hat{a}_{p,in}^{(1)}+\hat{a}_{s,in}^{(1)}$ with $\hat{a}_{s,in}^{(1)}$ being the intrinsic noise of the cavity. With Eq.~(\ref{eq:transducerlangevin}), the system operators can be derived as~\cite{MicrowaveOpticalTheo4}
\bea
\vec{v}(t)&=&U(t)e^{-i\int_{0}^{t}dt^{\prime}\Lambda(t^{\prime})}U^{-1}(0)\vec{v}(0)\nn\\
&+&\int_{0}^{t}dt^{\prime}U(t)e^{-i\int_{t^{\prime}}^{t}dt^{\prime\prime}\Lambda(t^{\prime\prime})}U^{-1}(t^{\prime})\sqrt{K}\vec{v}_{in}(t^{\prime}).\label{eq:sol}
\eea
Here $\Lambda(t)=\textrm{diag}(\lambda_{1},\lambda_{2},\lambda_{3})$ and $U(t) = [\psi_{1},\psi_{2},\psi_{3}]$, with $\lambda_{i}$ being the eigenvalues and $\psi_{i}$ the eigenmodes of the matrix $M(t)$. Assume no external input on cavity $\hat{a}_{2}$. The output photon of cavity $\hat{a}_{2}$ is $\hat{a}_{p,out}^{(2)}=\hat{a}_{p,in}^{(2)}-\sqrt{\kappa_{2,ext}}\hat{a}_{2}(t)$, where the input field $\hat{a}_{p,in}^{(2)}$ is in the vacuum state. In Fig.~\ref{fig4} (a), we plot the operator average of the output pulse $\ave{\hat{a}_{p,out}^{(2)}(t)}$ at two sets of parameters, in comparison to that of the input field. The time-domain solution gives us insight in understanding the relation between the input and the output signals~\cite{MicrowaveOpticalTheo3}.

The frequency dependence of the state conversion can be studied by transforming Eq.~(\ref{eq:transducerlangevin}) to the frequency domain. For constant couplings with $M(t)\equiv M_{0}$, the frequency component of the total output vector is $\vec{v}_{out}(\omega)=\widehat{T}(\omega)\vec{v}_{in}(\omega)$, where $\vec{v}_{out}(\omega) = [\hat{a}_{out}^{(1)}(\omega), \hat{b}_{out}(\omega), \hat{a}_{out}^{(2)}(\omega)]^{\textrm{T}}$ ~\cite{MicrowaveOpticalTheo2, MicrowaveOpticalTheo3, MicrowaveOpticalTheo4, MicrowaveOpticalExp3}, 
\be
\widehat{T}(\omega)=\left(I-i\sqrt{K}\left(I \omega-M_{0}\right)^{-1}\sqrt{K}\right)\label{eq:T}
\ee
is the conversion matrix, and $I$ is the identity matrix. The conversion matrix is a unitary matrix with $\widehat{T}^{-1}=\widehat{T}^{\dag}$. Note that the frequency $\omega$ is defined in the interaction picture. For the input (output) photon, $\omega=0$ corresponds to the frequency of cavity mode $\hat{a}_{1}$ ($\hat{a}_{2}$). The photon operator in the output of cavity $\hat{a}_{2}$ is
\bea
\hat{a}_{p,out}^{(2)}(\omega)=\hat{a}_{p,in}^{(2)}(\omega)+\sqrt{\frac{\kappa_{2,ext}}{\kappa_{2}}}\left[\widehat{T}_{31}(\omega)\hat{a}_{in}^{(1)}(\omega)\right.\nn\\
\left.+\widehat{T}_{32}(\omega)\hat{b}_{in}(\omega)+\lr{\widehat{T}_{33}(\omega)-1}\hat{a}_{in}^{(2)}(\omega)\right].\label{eq:aout2}
\eea
The conversion of quantum state from cavity $\hat{a}_{2}$ to cavity $\hat{a}_{1}$ can be derived similarly. The contribution of the quantum signal to the outgoing field is $\sqrt{\nu_{1}\nu_{2}}\widehat{T}_{31}(\omega)\hat{a}_{p,in}^{(1)}(\omega)$ with $\nu_{i}=\kappa_{i,ext}/\kappa_{i}$, which is implicitly included in $\hat{a}_{in}^{(1)}(\omega)$. The signal is reduced by a factor $\sqrt{\nu_{1}\nu_{2}}$ due to the intrinsic cavity dissipation. The matrix element $\widehat{T}_{31}(\omega)$, which links the input port of cavity $\hat{a}_{1}$ to the output port of cavity $\hat{a}_{2}$, determines the spectrum-dependence of the state conversion. In Fig.~\ref{fig4} (b), $\vert\widehat{T}_{31}(\omega)\vert$ is plotted for two sets of parameters. At $\omega=0$,
\be
\widehat{T}_{31}(0)=\frac{2\sqrt{\Gamma_{1}\Gamma_{2}}}{\Gamma_{1}+\Gamma_{2}+\gamma_{m}}.\label{eq:T310}
\ee
Here $\Gamma_{i}=4g_{i}^{2}/\kappa_{i}$ corresponds to the state conversion rate between an incoming photon and the mechanical mode as well as the cavity cooling rate. Under the impedance matching condition $\Gamma_{1}=\Gamma_{2}$ between the input and the output ports, $\widehat{T}_{31}(0)$ reaches a maximum with $\widehat{T}_{31}(0)\approx 1$ when $\Gamma_{i}\gg\gamma_{m}$. Another important property of the conversion matrix element is its halfwidth $\Delta\omega$, defined as $|\widehat{T}_{31}(\Delta\omega)|=|\widehat{T}_{31}(0)|/2$ ~\cite{MicrowaveOpticalTheo3}. In the strong coupling limit with $g_{i}\gg\kappa_{i}$, we find that $\Delta\omega\sim \kappa_{i}$ which is bounded by the cavity bandwidths. In the weak coupling limit with $g_{i}\ll\kappa_{i}$, $\Delta\omega\sim (\Gamma_{1}+\Gamma_{2}+\gamma_{m})$. Hence, for the parameter regime that is of interest to high-fidelity state conversion ($\Gamma_{i}\gg\gamma_{m}$ or cooperativity larger than $1$), the halfwidth is determined by the conversion rates $\Gamma_{i}$, as shown in Fig.~\ref{fig4} (b). For the opposite regime of $\Gamma_{i}\ll\gamma_{m}$, the halfwidth is limited by the mechanical damping rate $\gamma_{m}$. By designing optoeletromechanical systems with large cooperativity, the conversion bandwidth can surpass that of the mechanical resonance. An input state with a spectrum range within $\pm\Delta\omega$ of the cavity resonance could be converted with high efficiency. 

All other terms in Eq.~(\ref{eq:aout2}), including the vacuum fluctuations of $\hat{a}_{p,in}^{(2)}$, are noise terms. The mechanical noise makes the contribution $\sqrt{\nu_{2}}\widehat{T}_{32} \hat{b}_{in}(\omega)$ to the outgoing photon. Under the impedance matching condition, $\widehat{T}_{32}(\omega)/\widehat{T}_{31}(\omega)=i\sqrt{\gamma_{m}/\Gamma_{1}}$ at $\omega=0$, and does not increase significantly for a signal frequency within the conversion halfwidth $\Delta\omega$. The ratio of a quantum signal at the single-photon level to the mechanical noise is $\nu_{1}\sqrt{\Gamma_{i}/\gamma_{m}n_{th}^{(m)}}$. In Ref.~\cite{MicrowaveOpticalTheo4, TransducerOptics3DarkMode}, the role of the mechanical dark mode during the conversion of itinerant states was studied. It was found that under the impedance matching condition, the incoming field mainly excites the dark mode with the effect of the mechanical noise suppressed. The other noise contributions are $(-1)^{i+1}\sqrt{(1-\nu_{i})\nu_{2}}\hat{a}_{s,in}^{(i)}$ ($i=1,2$) and $(1-\nu_{2})\hat{a}_{p,in}^{(2)}$, after omitting the $\widehat{T}_{33}(0)$ terms~\cite{MicrowaveOpticalExp3}. These results indicate the importance of designing cavities with low intrinsic damping rates. 

Hence, to achieve high conversion efficiency for an itinerant quantum state at the single-photon level, the following conditions need to be satisfied. (1) The optomechanical interface is in the resolved-sideband regime with $\omega_{m}\gg\kappa_{i}$ so that the sidebands induced by the coupling are distinguishable from the cavity resonance and from each other. (2) The transducer satisfies the impedance matching condition which yields $\widehat{T}_{31}(0)\approx1$. (3) The thermal noise of the mechanical mode is bounded by $n_{th}^{(m)} < \Gamma_{i}/\gamma_{m}$ so that the magnitude of the quantum signal exceeds that of the noise terms in the output field. (4) The intrinsic cavity damping rate $\kappa_{i,in}$ is much weaker than the total damping rate $\kappa_{i}$ for both cavities. In current experiments, the conditions (1) and (2) can be readily fulfilled.
 
\section{Other applications: entanglement and quantum teleportation\label{sec:otherapp}}
The setup of an optoelectromechanical transducer can be utilized for other quantum applications, such as the generation of continuous-variable entanglement, squeezing, and quantum teleportation. These applications, together with the state conversion, enable various essential quantum information protocols to be performed by the transducer. An optoelectromechanical interface is hence a promising candidate for a tunable and noise-resilient hub in a hybrid quantum network. 

The generation of entanglement in electromechanical and optomechanical systems has been studied intensively over the past two decades, using methods such as stationary state scheme, transient approach, measurement based scheme, and quantum reservoir engineering~\cite{OptomechanicsReview1}. Here we focus on the ideas that are related to the quantum circuit in Sec.~\ref{subsec:transducer}~\cite{Entanglement2Mode1, Entanglement2Mode2, Entanglement2Mode3, Entanglement3Mode1, Entanglement3Mode2}. Let cavity $\hat{a}_{1}$ be driven by a red-detuned field with the detuning $\Delta_{1}=-\omega_{m}$ and cavity $\hat{a}_{2}$ be driven by a blue-detuned field with the detuning $\Delta_{2}=\omega_{m}$, as illustrated in Fig.~\ref{fig2} (d). The effective coupling is $H_{I}=(\hbar g_{1} a_{1}^{\dagger}b_{m}+i\hbar g_{2}a_{2}^{\dagger}b_{m}^{\dagger}+h.c.)$~\cite{EntanglementInterface1, EntanglementInterface2, EntanglementInterface3, EntanglementInterface4, EntanglementInterface5}. The blue-detuned driving induces a parametric-down-conversion operation between cavity $\hat{a}_{2}$ and the mechanical mode, which is a key ingredient in entanglement generation. This operation alone, however, induces instability in the system and cannot produce very strong entanglement~\cite{Entanglement2Mode1}. Adding the red-detuned driving on cavity $\hat{a}_{1}$ (the beam-splitter-like operation) serves to balance the effect of the blue-detuned driving, and makes the system stable under the condition $g_{1}>g_{2}$. The red-detuned driving also removes thermal fluctuations from the mechanical mode. It was shown that strong entanglement robust against the thermal noise can be generated between intra-cavity photons at selected time windows as well as between output photons near the frequency $\omega=0$. This can be explained by the notion of the Bogoliubov dark mode $\hat{\beta}= \cosh(r)\hat{a}_{2}+i\sinh(r)\hat{a}_{1}^{\dagger}$ with $r=\tanh^{-1}(g_{2}/g_{1})$ and $g_{0}=\sqrt{g_{1}^{2}-g_{2}^{2}}$. This mode is an eigenmode of the interaction Hamiltonian $H_{I}$ with eigenvalue zero and does not contain the mechanical component. The noise-resiliency of the entanglement is a combination of the excitation of this dark mode and the quantum interference between the bright modes~\cite{EntanglementInterface3}. The mechanical noise can be eliminated to leading order. This scheme was analyzed in the picture of quantum reservoir engineering~\cite{EntanglementInterface2, EntanglementInterface4}, where the dark mode corresponds to a two-photon squeezed vacuum state. This idea was also extended to the generation of entanglement between two mechanical modes coupling to one cavity mode~\cite{EntanglementInterface2, EntanglementInterface4}. 
 
In another work, a S\o renson-M\o lmer-like approach originally developed for implementing quantum logic gates on trapped ion qubits was studied to generate robust entanglement between two cavity modes~\cite{EntanglementInterface5}. The detunings of the pump fields are adjusted at selected times to achieve an effective two-mode squeezing operation on the cavity modes. It was shown that the entanglement can survive high bath temperature and can be implemented in the weak-coupling limit.

\begin{figure}
\includegraphics[width=\columnwidth,clip]{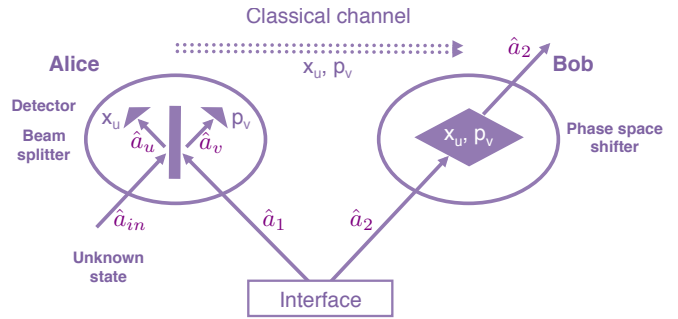}%
\caption{Quantum teleportation scheme~\cite{TeleportationInterface2}. The entangled photons $\hat{a}_{1}$ and $\hat{a}_{2}$ are sent to Alice and Bob, where local operations are performed. The unknown microwave (optical) state $\hat{a}_{in}$ is teleported to an optical (microwave) photon $\hat{a}_{2}$.}
\label{fig5}
\end{figure}

Given the entangled photons, quantum teleportation of unknown states can be realized in an hybrid optoelectromechanical interface, as shown Fig.~\ref{fig5}~\cite{TeleportationInterface2}. The entangled photons are delivered to distant nodes that have no direct interaction. The unknown quantum state is initially stored in one of the nodes. Local beam-splitter operation followed by Bell-state measurements is performed on the unknown state and one of the photons in the entangled pair. Afterwards, the measurement results are transferred to the other node via classical communication channels. The classical information is used to shift the state of the other photon in the entangled pair, which then recovers the unknown state. With this setup, the unknown state can be teleported from microwave to optical wavelength, or vice versa~\cite{ContinuousVariableRMP}.

\section{Experimental progress\label{sec:experiments}}
\subsection{Resolved-sideband regime and coherent effects\label{subsec:coherence}}
The past few years have witnessed numerous experimental advances in the electromechanical and optomechanical systems. Here we summarize the key experiments that paved the way for realizing coherent conversion of photons via an optoelectromechanical interface. The resolved-sideband regime with $\omega_{m}\gg\kappa$, a necessary condition for implementing the quantum applications discussed in the previous sections, was demonstrated in both microwave and optical cavities~\cite{ResolvedSideband1, ResolvedSideband2, ResolvedSideband3}. For a superconducting microwave resonator with $\omega_{c}/2\pi=10\,\textrm{GHz}$ and a moderate quality factor of $Q=10^{4-5}$, this regime can be easily accessed for a mechanical mode of $\omega_{m}/2\pi\sim10\,\textrm{MHz}$~\cite{Groundstate3}. In the optical domain, reaching this regime demands cavity modes with very high quality factor~\cite{strongcouplingExp4} and mechanical modes with relatively high frequency~\cite{Groundstate5, MechanicalInterface6}. In the resolved-sideband regime, ground state cooling of the mechanical mode via quantum backation was realized~\cite{Groundstate1, Groundstate2, Groundstate3, Groundstate4, Groundstate5, Groundstate6, Groundstate7}. The effective linear coupling between the cavity and the mechanical modes can be adjusted from the weak coupling limit ($g\ll \kappa$) to the strong coupling limit ($g\gg\kappa$) by varying the driving power. In the weak coupling limit, the mechanically-induced transparency was observed in the transmission spectrum of a probe field in a number of systems, including an optomechanical system in toroidal microcavity, an optomechanical crystal structure, a microwave resonator coupling with a membrane, and a membrane-in-the-middle setup~\cite{EITExp1, EITExp2, EITExp3, EITExp4, MicrowaveOpticalExp1, MicrowaveOpticalExp2}. In the strong coupling limit, the normal-mode spitting was shown in the transmission spectrum, which indicates the excitation of polariton-like eigenmodes of the optomechanical system~\cite{strongcouplingExp3, strongcouplingExp4, strongcouplingExp5, strongcouplingExp6, strongcouplingExp7, MicrowaveOpticalExp2}. Driving the cavity mode with a detuned field also results in the amplification of the cavity and the mechanical signals, which was demonstrated in the optical regime in Ref.~\cite{Amplification1}. Meanwhile, in a recent experiment, a microwave resonator coupling to a mechanical mode is driven by a blue-detuned source that generates a parametric-down-conversion coupling between the cavity and the mechanical modes~\cite{Amplification2}. Mechanical microwave amplification of $25$ decibels was achieved on a probe field near the cavity resonance due to the stimulated emission induced by the probe field. 

\subsection{Converting photon to mechanical memory\label{subsec:storage}}
Quantum states stored in a mechanical mode could be preserved for a duration of $\sim1/\gamma_{m}n_{th}$. Recently, experiments were performed to transfer itinerant photons to a mechanical mode that serves as a quantum memory for the photon states~\cite{TranferMechanicalMode1, TranferMechanicalMode2}. These experiments manifested the electromechanical or optomechanical system as an interface to connect photons to the mechanical mode.  

In an experiment performed on a deformed silica microsphere~\cite{TranferMechanicalMode1}, optical information was converted to and from a mechanical excitation. The mechanical mode is a breathing mode of the microsphere with a frequency of $\omega_{m}/2\pi=108\,\textrm{MHz}$ and the optical mode is a whispering gallery mode with $\kappa/2\pi=40\,\textrm{MHz}$. The writing and readout of the mechanical mode were performed by applying red-detuned pulses (as in the left panel of Fig.~\ref{fig1}) separated by a finite delay time. The effective coupling strength is $\sim 2\,\textrm{MHz}$, which is in the weak coupling limit. During the writing pulse, an optical field at the cavity frequency is injected to the resonator via free-space evanescent excitation, which is converted to a mechanical state with a coupling efficiency of $9\%$. The dynamics of this process is described by Eq.~(\ref{eq:mconvert}) under the RWA. During the readout process, the mechanical state is converted back to a propagating optical signal. The power of the retrieved pulse decays exponentially with the delay time due to the damping of the mechanical mode. The storage life time is $\sim 3.5\,\mu\textrm{s.}$. This experiment provides a proof-of-principle demonstration of the state transfer protocol and the mechanical storage of the light field. 

The coherent conversion of a microwave photon to a mechanical state was implemented with a microwave resonator capacitively coupled to a membrane~\cite{TranferMechanicalMode2}. The superconducting aluminium membrane forms the upper plate of a capacitor in the microwave resonator. The electromechanical interaction originates from the modification of the capacitance by the vibration of the membrane, as illustrated in Fig~\ref{fig2} (a). The frequency of the microwave resonator is $\omega_{c}/2\pi=7.5\,\textrm{GHz}$ with the bandwidth $\kappa/2\pi\sim 320\,\textrm{kHz}$ and the frequency of the drumhead mode of the membrane is $\omega_{m}/2\pi=10.5\,\textrm{MHz}$. The transfer and the signal pulses are coupled inductively to the microwave resonator by a coplanar-waveguide transmission line. The itinerant microwave signal has the pulse shape $V_{in}(t)\propto \exp{(\Gamma_{prep}t/2)}\Theta(-t)$ with $\Theta(t)$ being the unit step function. During a delay time $\tau_{del}$, the converted state is stored in the mechanical mode. A second transfer pulse is then applied to retrieve the mechanical state to an outgoing pulse. The experiment demonstrated that optimal transfer of the incoming field to the mechanical state is achieved at $\Gamma_{prep}\approx\Gamma$ in the limit of weak coupling. The phase and amplitude of the incoming field are preserved in the mechanical state. The variances of the mechanical state are charactered in the quadrature phase space by considering the quantum efficiency of the measurement and of the state transfer process (less than unity due to the intrinsic resonator damping). The time-dependence of the heating of the mechanical mode is inferred from the measurement. By applying a strong transfer pulse, the mechanical state can be swapped to an intra-cavity state of the resonator with an oscillatory dependence on the pulse length. This work opens the possibility of state transfer in the quantum limit. Meanwhile, in a related work, this device was utilized to generate entanglement between propagating microwave photon and the mechanical state by applying a blue-detuned driving on the resonator~
\cite{Entanglement2Mode3}.

\begin{figure}
\includegraphics[width=\columnwidth,clip]{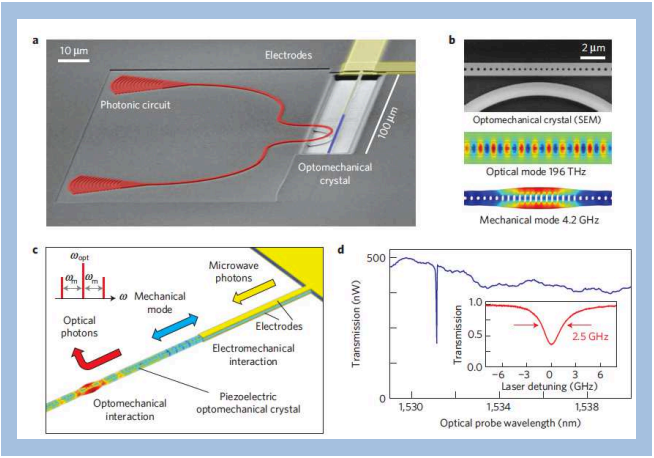}%
\caption{An optoelectromechanical transducer made of a piezoelectric optomechanical crystal. The external electrodes connect the transducer to the electrical driving. Adapted by permission from Macmillan Publishers Ltd: Ref.\cite{MicrowaveOpticalExp1}, \textcopyright~2013.}
\label{fig6}
\end{figure}
\subsection{Converting microwave to optical photons\label{subsec:conversion}}
Several experimental setups for an optoelectromechanical transducer have been studied for converting microwave to optical photons. In these experiments, a microwave field is injected to the transducer either by propagating through a microwave cavity or by directly exciting the mechanical mode. A schematic circuit of a transducer made of a piezoelectric optomechanical crystal is shown in Fig.~\ref{fig6}~\cite{MicrowaveOpticalExp1}. In this setup, a mechanical breathing mode with a microwave frequency of $\omega_{m}/2\pi=4\,\textrm{GHz}$ and an optical cavity with a frequency of $196\,\textrm{THz}$ and a bandwidth of $\kappa/2\pi=2.5\,\textrm{GHz}$ are co-designed in a photonic bandgap structure, coupling to each other via optomechanical interaction. The device is operated at room temperature. The microwave field couples capacitively to the mechanical resonator and generates a mechanical excitation with average phonon number $N_{exc}\approx10^{6}$ via piezoelectric actuation. For an optical field injected to the transducer from the external circuit, the parametric optomechanical coupling induces optical sidebands. Homodyne detection of the sidebands reveals the frequency-modulation nature of the coupling. The detected signal maintains the phase and amplitude information of the microwave field, indicating the coherent transfer of the microwave field to an optical signal. The experiment was also conducted for reduced driving magnitude with small excitation numbers $N_{exc}=100,\,7,\,1$ at a thermal background of $n_{th}^{(m)}\approx 1500$. A narrow transparency window of the optical probe appears in the presence of the mechanical excitation when the probe detuning $\Delta_{p}$ equals the frequency of the mechanical excitation $\Omega_{exc}$. This effect is an interference effect, similar to the mechanically-induced transparency discussed in Sec.~\ref{subsec:coherenceeffect}. With the mechanical frequency in the microwave range, this device facilitates the direct coupling of microwave photons to the optomechanical interface~\cite{MechanicalInterface6}. By making the system compatible with cryogenic environment, it would be possible to connect the mechanical resonator to superconducting qubits and resonators to form a hybrid quantum network. 

\begin{figure}
\includegraphics[width=\columnwidth,clip]{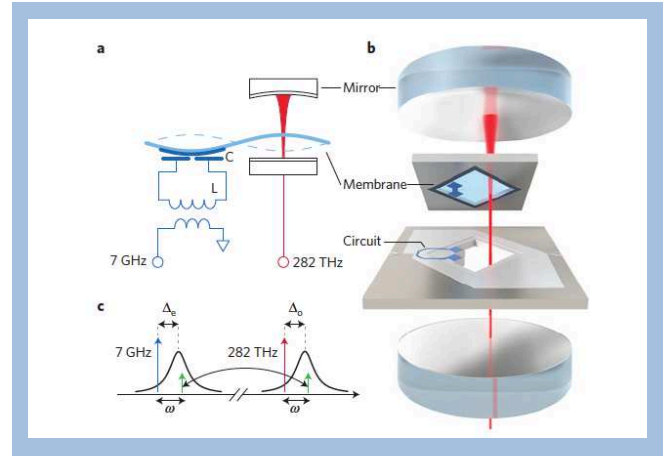}%
\caption{An optoelectromechanical transducer formed by a microwave resonator and an optical cavity both coupling to a membrane. Adapted by permission from Macmillan Publishers Ltd: Ref.~\cite{MicrowaveOpticalExp3}, \textcopyright~2014.}
\label{fig7}
\end{figure}
Coherent bi-directional state conversion between microwave and optical photons was recently demonstrated in a hybrid system made of a microwave resonator and an optical cavity coupling to a membrane~\cite{MicrowaveOpticalExp3}. This transducer has the same configuration as discussed in Sec.~\ref{subsec:itinerant} and is operated at a temperature of $4\,\textrm{K}$. One part of the membrane couples capacitively to a microwave resonator with a frequency of $7\,\textrm{GHz}$, and another part of the membrane couples to a Fabry P\'{e}rot cavity with a frequency of $282\,\textrm{THz}$, as shown in Fig.~\ref{fig7}. Each cavity mode is driven by a strong red-detuned pump field which generates tunable coupling between the cavity and the mechanical modes. The conversion gain is $A=A_{e}A_{o}$ with $A_{i}=1+(\kappa_{i}/\omega_{m})^{2}$ for the cavities. Quantum state conversion requires unit gain $A_{i}=1$ with the system in the resolved-sideband regime. For a conversion performed on a mechanical mode of the frequency $\omega_{m}/2\pi=560\,\textrm{kHz}$, $A_{i}\approx\sqrt{1.4}$ at the cavity bandwidths of this device. An electric (optical) signal can be coupled to the microwave (optical) port, and converted to the optical (microwave) port. The frequency dependence of the state conversion as well as the correlation of the conversions along both directions were characterized. The experiment shows a conversion efficiency of $\sim10\%$, far surpassing previous results in the microwave-to-optical domain~\cite{ElectroOptics1}. This efficiency is limited by intrinsic cavity dissipation and optical mode mismatching. It was predicted that when operated at a low temperature of $40\,\textrm{mK}$, quantum state conversion in the single-photon level could be achieved in this transducer. 

In another experiment~\cite{MicrowaveOpticalExp2}, the transmission of a radio-frequency signal to an optical field was realized in a room-temperature optoelectromechanical transducer. Here a nanomembrane is part of a capacitor in a radio-frequency resonance circuit. The frequency of the resonance circuit is adjusted to be near the frequency of the fundamental drum mode of the membrane with $\omega_{m}=0.72\,\textrm{MHz}$. The bandwidth of the resonance circuit is $5.5\,\textrm{kHz}$, deeply in the resolved-sideband regime. The electromechanical coupling between the electrical mode and the mechanical mode is proportional to and tunable by a static bias voltage. The coupling is in the strong coupling limit with high cooperativity at a bias voltage of $10$ volts. A radio-frequency signal inductively coupled to the resonance circuit excites a voltage modulation in the circuit and a displacement modulation of the membrane. Optical light that is reflected off the surface of the membrane carriers optical sidebands induced by the mechanical motion, which is a phase-insensitive transduction of the electrical signal to the optical field. Measurement of the optical light can hence be used as a readout of the mechanical motion. Mechanically-induced transparency and normal-mode splitting are observed in both the electrical output port and the reflected optical signal. 

State conversion in the optical-to-optical domain was performed on optomechanical interfaces between photons of distinct frequencies. As we mentioned in Sec.~\ref{subsec:itinerant}, when the impedance matching condition is fulfilled, the probe pulse is converted mostly via the mechanical dark mode, and the effect of the mechanical noise is suppressed. In Ref.~\cite{TransducerOptics3DarkMode}, the dark mode was studied with optical whispering gallery modes and a mechanical breathing mode in a deformed silica sphere. The dark mode is excited when injecting an optical probe pulse at the resonance of one cavity mode. By adjusting the strengths of the optomechanical couplings $g_{1,2}$, the cavity components in the dark mode are varied and the photon emission from the cavities is affected accordingly. This experiment verifies the role of the mechanical dark mode in the transmission. In another experiment, wavelength conversion between telecommunications frequencies was demonstrated with a transducer made of silicon optomechanical crystal nanobeam~\cite{TransducerOptics3}. The mechanical mode in this experiment has a frequency of $4\,\textrm{GHz}$. An optical signal can be converted between the wavelengths of $1146\,\textrm{nm}$ and $1545\,\textrm{nm}$ with a high internal efficiency of $93\%$. These experiments verified the possibility of achieving high conversion efficiency via mechanical resonators. 

\section{Challenges and perspectives\label{sec:discussions}}
In Sec.~\ref{subsec:itinerant}, we discussed the requirements for achieving high efficiency state conversion via an optoelectromechanical transducer. Given the recent experimental progress, both microwave and optical systems can reach the resolved-sideband regime either by improving the cavity quality factor or increasing the mechanical frequency~\cite{MechanicalInterface6}. The impedance matching condition can be satisfied by tuning the effective electromechanical and optomechanical couplings. Note that in the sideband-unresolved regime with $\kappa>\omega_{m}$, the conversion gain can be larger than unity which increases the conversion efficiency. But the transducer becomes a phase-insensitive amplifier, adding extra noise to the output state and destroying the quantum coherence~\cite{MicrowaveOpticalExp3}. The major obstacles that prevent the realization of quantum state conversion include the perturbation of the mechanical noise and the intrinsic dissipation of the cavity modes. To distinguish a single-photon level signal in the cavity output, it requires $4g_{i}^{2}/\kappa_{i}\gamma_{m}> n_{th}^{(m)}$, which puts an upper bound on the thermal occupation number of the mechanical mode. This condition can be satisfied by introducing cryogenic setups to lower the operating temperature of the transducer or by designing samples with high cooperativity. The intrinsic cavity dissipation not only presents another source of noise in the output state, but also reduces the magnitude of the signal. To obtain high conversion efficiency, it requires that $\kappa_{i,in}\ll \kappa_{i,ext}$ with negligible internal loss in the cavities. This puts a restriction on the driving power that can be applied to the cavities, which could generate excess heating, and on the designs of the circuit. Besides the noise sources presented in the transducer, it is also challenging to overcome the technical difficulties in connecting the microwave and the optical devices with a mechanical resonator, while keeping the devices functioning in the quantum limit. 

In summary, recent experiments have demonstrated that an optoelectromechanical transducer can bridge the microwave and the optical worlds and enable coherent state conversion between microwave and optical photons. These experimental advances manifest that high efficiency quantum state conversion at the single-photon level could be achieved in near future. Given the diversity of the electromechanical and optomechanical systems, various experimental setups can be studied to construct such a hybrid quantum interface. Besides the photon conversion protocols, the transducer circuit can also be used for the conversion between qubit and photon states or between intra-cavity and propagating photons. An optoelectromechanical interface can hence serve as a tunable and noise-resilient hub in a hybrid quantum network to perform various quantum information protocols such as quantum wavelength conversion and quantum teleportation. 

\begin{acknowledgements}
We thank Prof. Konrad W. Lehnert for careful reading of the manuscript and for very helpful comments. This work is supported by the DARPA ORCHID program through AFOSR, the National Science Foundation under Award Numbers NSF-DMR-0956064 and NSF-CCF-0916303, and the NSF-COINS program under Grants No. NSF EEC-0832819.
\end{acknowledgements}

\end{document}